\documentclass[prl,aps,twocolumn,showkeys,showpacs,superscriptaddress]{revtex4-2} % 
\usepackage{graphicx} % needed for figures
\usepackage{dcolumn} % needed for some tables
\usepackage{bm}  % for math
\usepackage{amssymb} % for math
\usepackage{amsmath} % for math
\usepackage{xcolor}

\usepackage{booktabs}
\usepackage{hyperref}
\usepackage{cancel}
\begin{document}
\title{Elastic phase shift analysis reveals the geometric origin of the residue phase}

\author{S.~Ceci}
\email{sasa.ceci@irb.hr}
\affiliation{Rudjer Bo\v{s}kovi\'{c} Institute, Bijeni\v{c}ka 54, HR-10000 Zagreb, Croatia}
\author{R.~Omerovi\'{c}} 
\affiliation{University of Tuzla, Urfeta Vejzagi\'{c}a 4, 75000 Tuzla, Bosnia and Herzegovina}
\author{H.~Osmanovi\'{c}} 
\affiliation{University of Tuzla, Urfeta Vejzagi\'{c}a 4, 75000 Tuzla, Bosnia and Herzegovina}
\author{M.~Uroi\'{c}}
\affiliation{Rudjer Bo\v{s}kovi\'{c} Institute, Bijeni\v{c}ka 54, HR-10000 Zagreb, Croatia}
\author{M.~Vuk\v{s}i\'{c}}
\affiliation{University of Oxford, Oxford OX1 2JD, United Kingdom}
\author{B.~Zauner}
\affiliation{Institute for Medical Research and Occupational Health, Ksaverska 2, HR-10000 Zagreb, Croatia}
\date{\today}

\begin{abstract}
We show that the complex-plane structure of light hadron resonances is governed by a unified geometric framework where the threshold position plays a decisive role. By applying this framework to $\pi\pi$, $\pi K$, and $\pi N$ phase shifts, we show that the residue phase $\theta$ is primarily determined by the geometric phase $\delta_0$ (the angle between pole and real axis seen from the threshold). While vector resonances exhibit excellent alignment with this geometric baseline, scalar resonances show systematic deviations of $10^\circ$--$15^\circ$, which we identify as the dynamical imprint of Adler zeros.
\end{abstract}

\keywords{Light non-strange mesons, Light strange mesons, Nucleon resonances, Analytic properties of S matrix}
\pacs{14.40.Be, 14.40.Df, 14.20.Gk, 11.55.Bq}

\maketitle

%\section{Introduction}

{\bf Introduction.}---Resonances are fundamentally defined as simple poles of the S-matrix on non-physical Riemann sheets \cite{DalitzMoorhouse}. While the real and imaginary parts of the pole position are the resonance mass $M$ and total width $\Gamma$, and the residue magnitude defines the coupling strength, the physical origin of the residue phase $\theta$ has remained elusive \cite{PDG}. Traditionally attributed to background interference and coupling phases, recent findings suggest that for the lightest resonances, S-matrix unitarity constrains $\theta$ to a fundamental geometric identity \cite{Cec26,Cec26B}.

In this Letter, we use this constraint to resolve the complex-plane structure of the lightest scalar and vector mesons. We systematically contrast the broad, non-canonical scalar states---the $f_0(500)$ in $\pi \pi$ scattering and $K_0^*(700)$ in $\pi K$ scattering---with their canonical vector counterparts, the $\rho(770)$ and $K^*(892)$. While the unusual scattering line shapes of the $f_0(500)$ and $K_0^*(700)$ are conventionally attributed to strong backgrounds and subthreshold Adler zeros \cite{Adl65}, we demonstrate that their complex-pole properties, along with the narrow $\Delta(1232)$ nucleon resonance, are fundamentally unified through a single, mathematically rigid threshold geometry.

%\section{Model}

{\bf Theoretical Framework.}---We adopt the single resonance scattering matrix parametrization from Refs.~\cite{Cec26,Cec26B}, which is valid in the proximity of the resonance pole, assume an elastic resonance and impose S-matrix unitarity. In that case, the complex phase of the scattering amplitude is given by the phase shift
\begin{equation}
    \delta = \arctan\frac{\Gamma/2}{M-E} + \delta_\mathrm{B}, \label{eq:phaseshift}
\end{equation}
with the approximately constant background term $\delta_\mathrm{B}$. Here, $E$ is center of mass energy. 

Up to this point, this may seem like the textbook Breit-Wigner parameterization. However, the key contribution from Refs.~\cite{Cec26,Cec26B} is that for elastic resonances $\delta_\mathrm{B} = \delta_0$, where the geometric phase $\delta_0$ is defined as
\begin{equation}
    \delta_0= -\arctan\frac{\Gamma/2}{M-E_0}.
\end{equation}
Here, $E_0$ is the elastic threshold energy. The geometrical meaning of the $\delta_0$ is the angle between the real axis and the pole position seen from the threshold. 

This framework assumes that the background phase $\delta_\mathrm{B}$, as well as the pole parameters $M$ and $\Gamma$, are locally constant. While $\Gamma$ and $\delta_\mathrm{B}$ are naturally energy-dependent in fully dynamical models, locally assuming their constancy yields the exact elastic residue phase $\theta=\delta_\mathrm{B}$ \cite{Hoehler}.

\begin{figure*}
\includegraphics[width=0.24\textwidth]{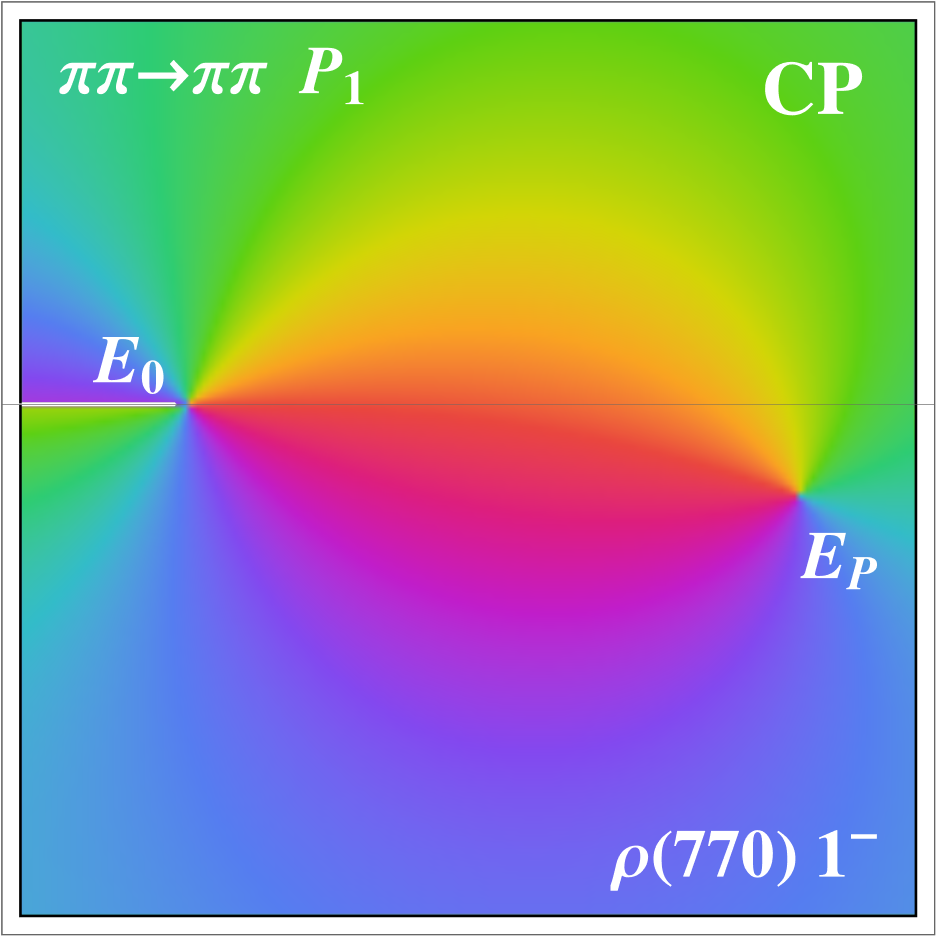}
\includegraphics[width=0.24\textwidth]{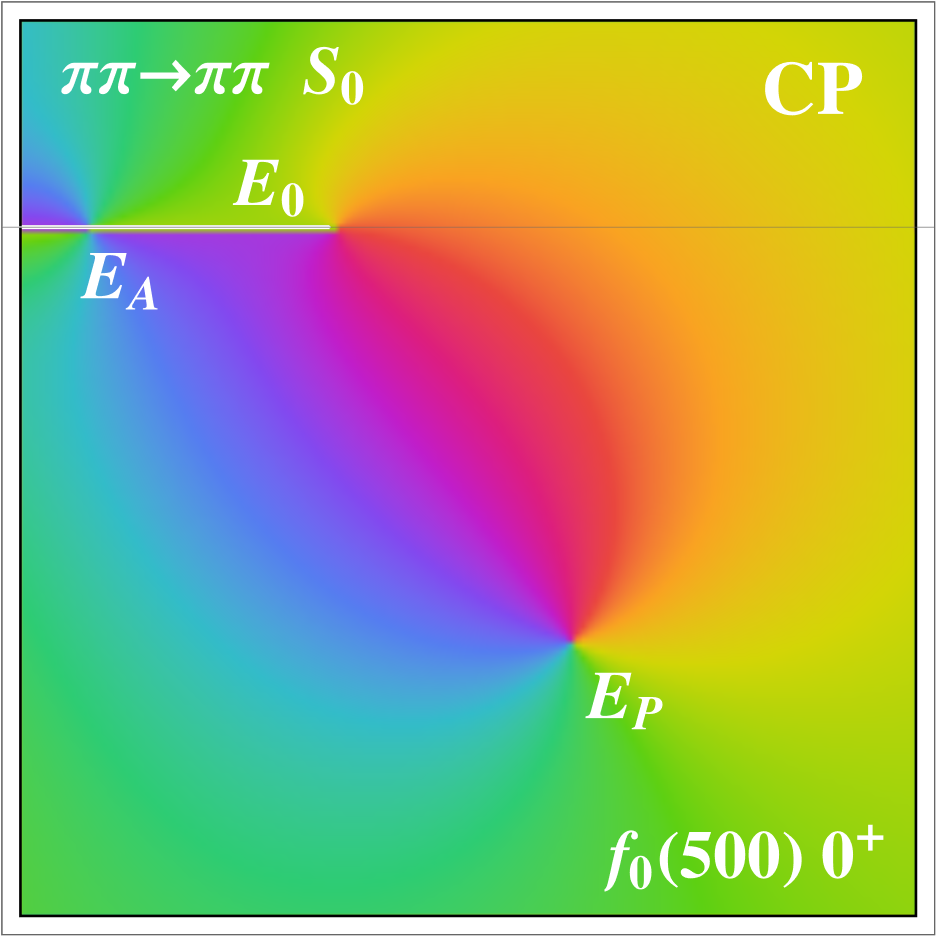}
\includegraphics[width=0.24\textwidth]{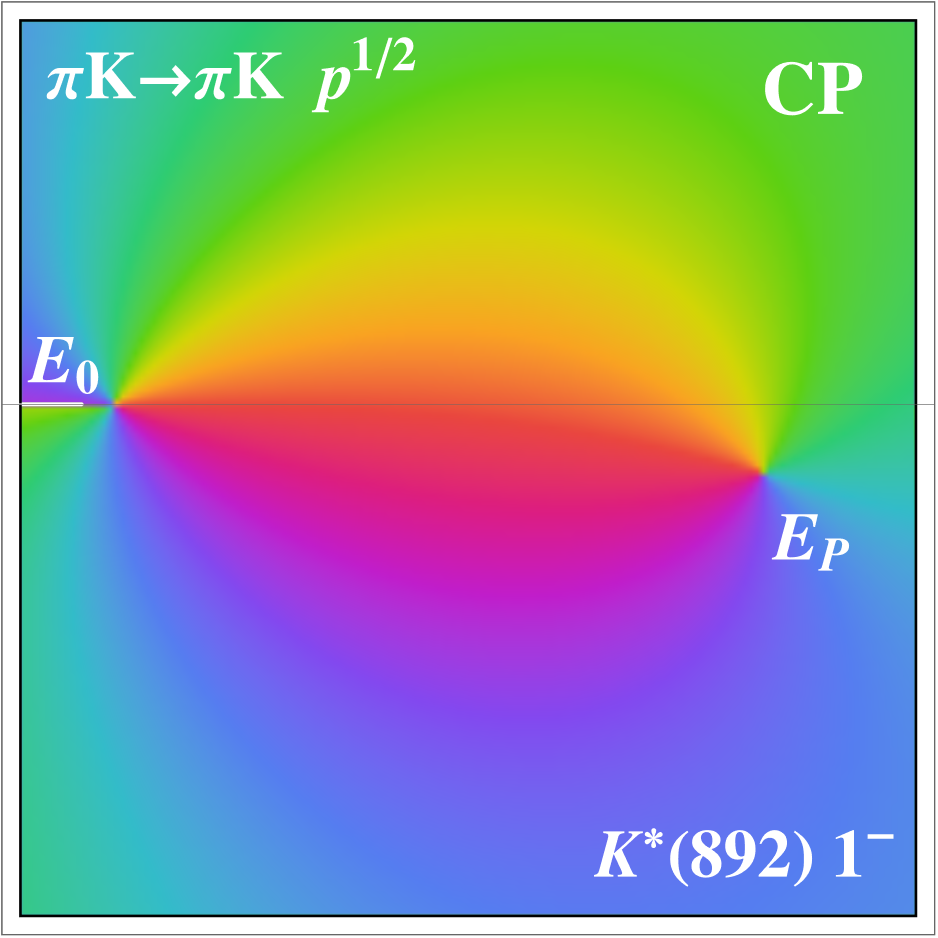}
\includegraphics[width=0.24\textwidth]{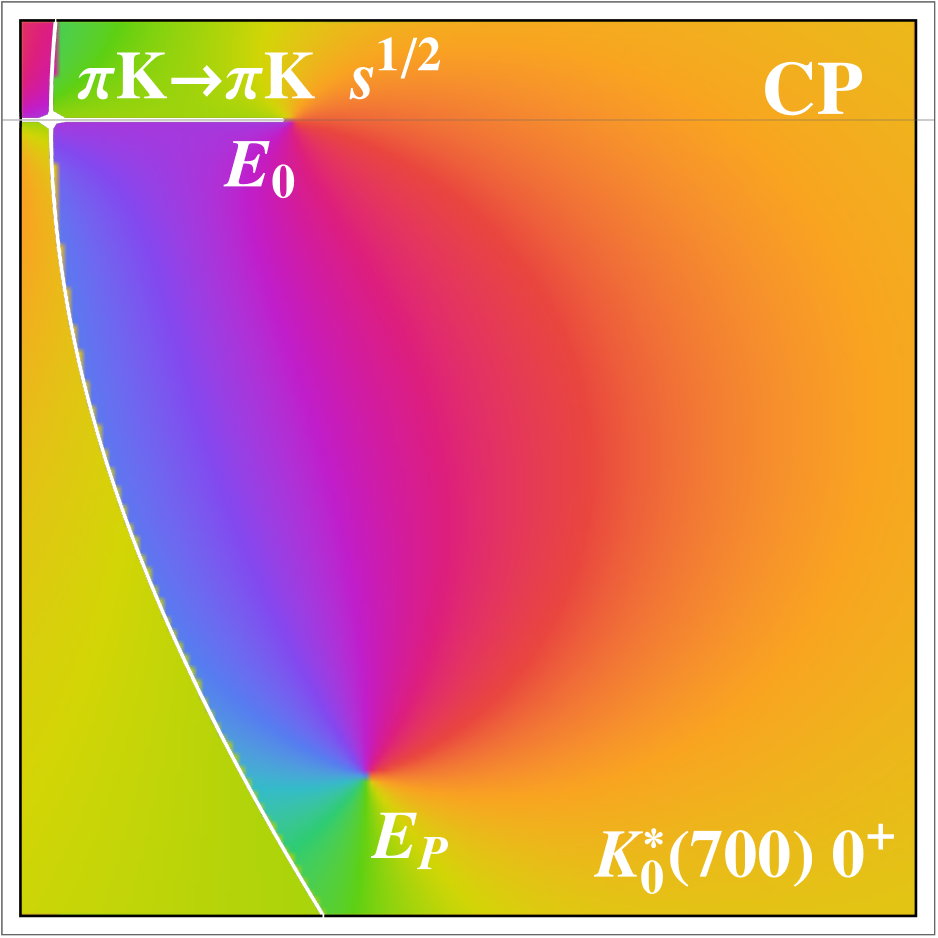}
\includegraphics[width=0.24\textwidth]{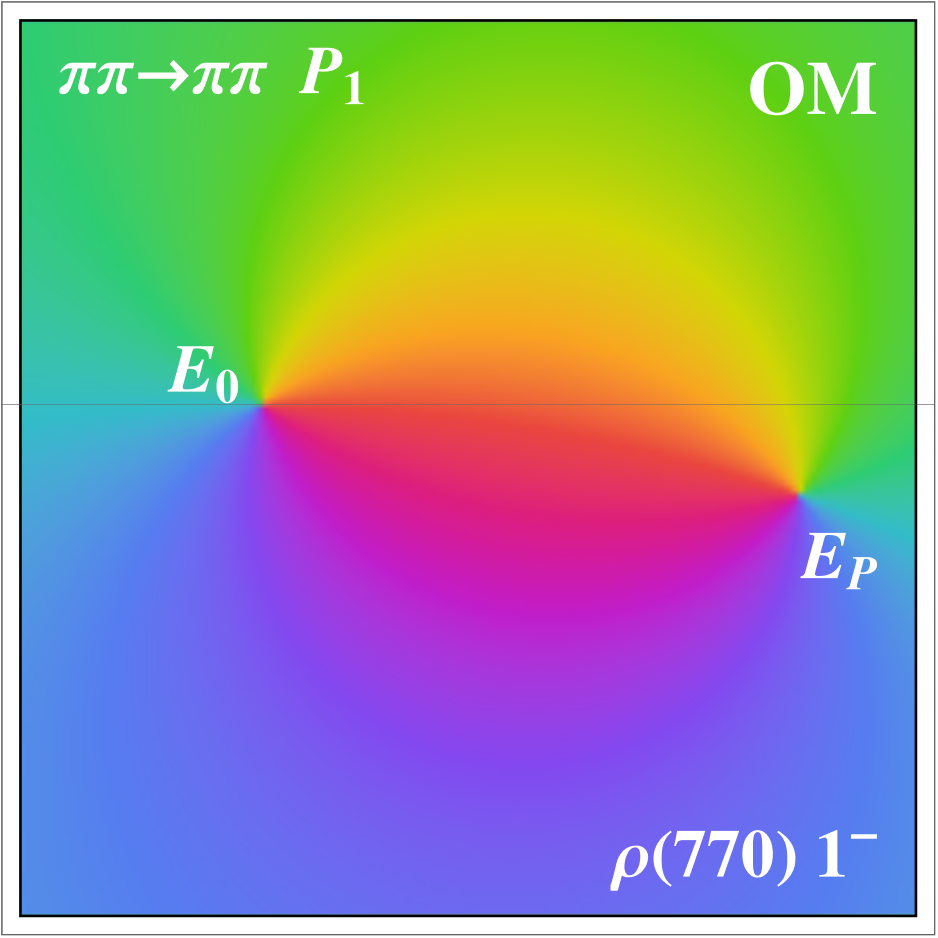}
\includegraphics[width=0.24\textwidth]{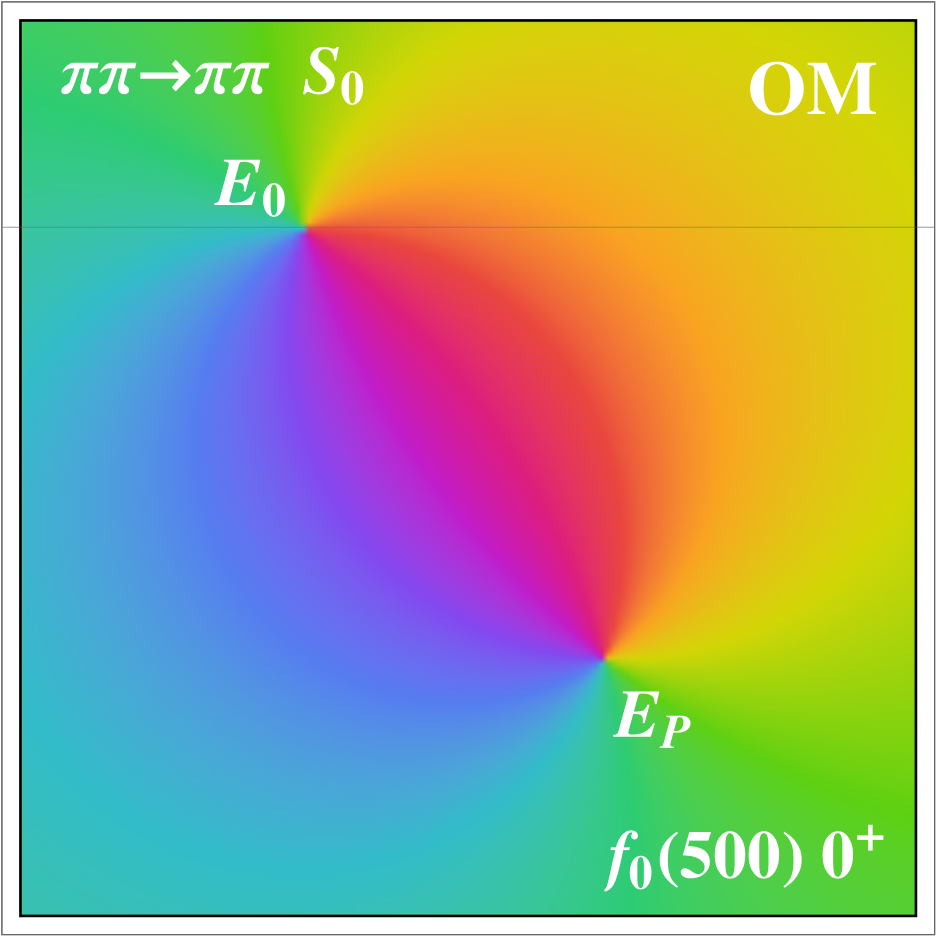}
\includegraphics[width=0.24\textwidth]{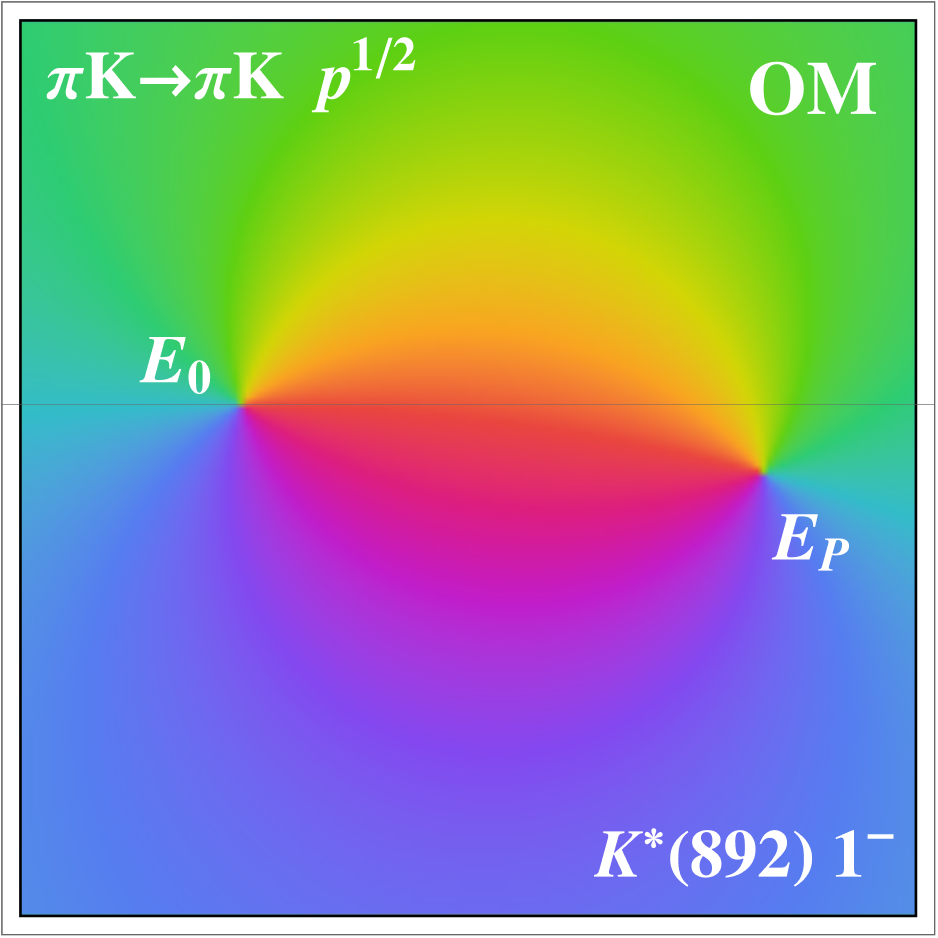}
\includegraphics[width=0.24\textwidth]{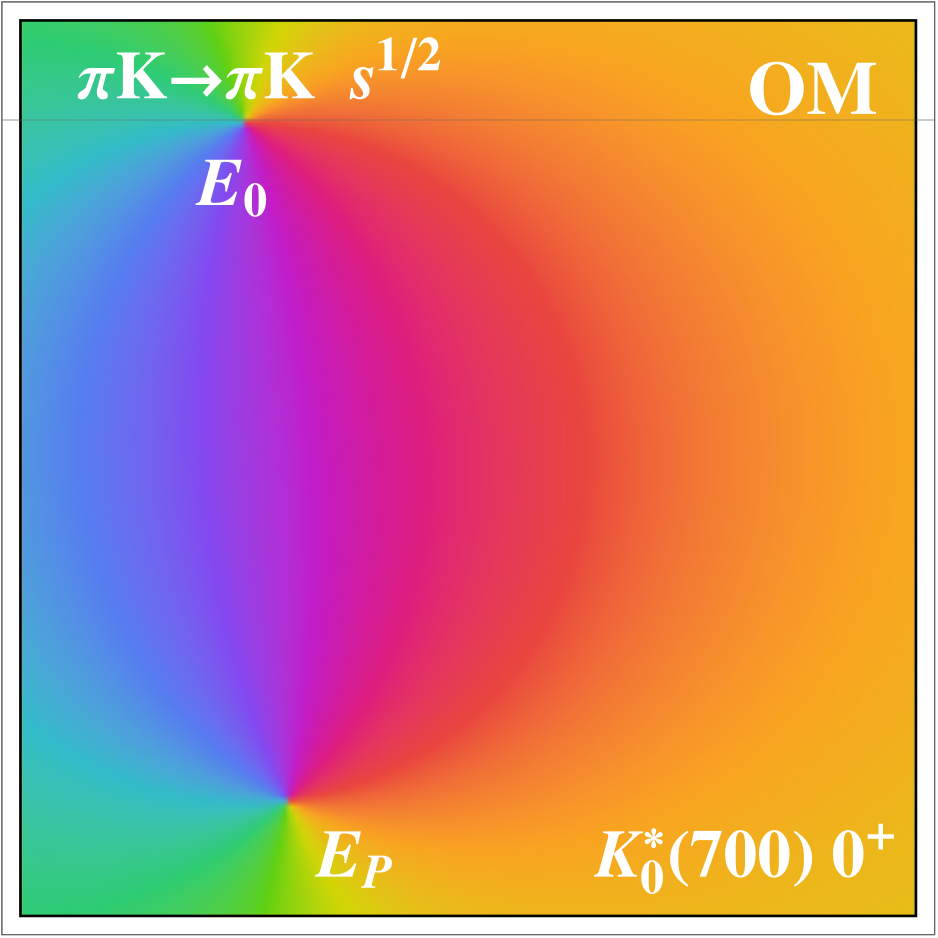}
\caption{ Scattering amplitude phase ($\arg \,T$) in the complex energy plane. (Upper half) Amplitude is built from naturally expanded theoretical $\delta$ from Refs.~\cite{PR,GM} into the complex energy plane (CP). Here, $E_0$ is the threshold and $E_P$ is the pole position. Adler zero ($E_A$) is clearly seen for $f_0(500)$, while for $K_0^*(700)$ it is obscured by the conformal variable cut structure. (Lower half) Amplitude constructed from the fitted $\delta$ in our model (OM). The poles $E_P$ are clearly visible, while the thresholds $E_0$ are effectively described as simple zeros. 
\label{Fig1}}
\end{figure*}

To test our hypotheses, we generate theoretical data points for $\delta$ using the parametrizations from Refs.~\cite{GM,PR,Arndt}. We search for the poles and residues of scattering amplitude and we reconstruct it in the complex plane (CP) in the upper half of Fig.~\ref{Fig1}, using relation $T=(\cot\delta-i)^{-1}$.

{\bf Phase Conversion.}---A persistent challenge in evaluating resonance parameters is the apparent discrepancy between the unitary residue phase $\theta$ extracted in the complex-energy plane and the bare coupling phase $\phi_g$ reported by high-precision dispersive analyses \cite{GM,PR,Hof24}. By strictly adopting the baryon resonance convention for the residue limit, $\lim_{E\to E_p}(E_p-E)\,T(E)$, we show that this discrepancy is fundamentally a kinematic artifact.

The phase $\theta$ is universally linked to the phase $\phi_g$ via the exact relation:
\begin{equation}
    \theta = 2\phi_g - 2\arg(E_p) + (2l+1)\arg(q(E_p)),\label{eq:Convention}
\end{equation}
where $q(E_p)$ is the momentum evaluated at the pole energy $E_p$, and $l$ is the angular momentum number.

This translation absorbs both the Jacobian rotation from the Mandelstam s-plane to the E-plane and the complex centrifugal barrier kinematics. When this translation is applied, the apparent offset vanishes, and our geometrically derived phases exhibit striking agreement with results from the literature (cf.~Table \ref{Table1}).

{\bf Higher-Order Derivative Diagnostics.}---To distinguish genuine resonance poles from dynamically generated structures without relying on deep complex-plane extrapolation (CP), we introduce a diagnostic test based on higher-order energy derivatives of the phase shift~\cite{Cec08}. Extending the traditional first-derivative time-delay methods \cite{Wigner, Hoehler}, we exploit the property that for a generic Breit-Wigner resonance with a locally constant background $\delta_\mathrm{B}$, all even-order energy derivatives $\delta^{(2n)}(E)$ of Eq.~\ref{eq:phaseshift} must cross zero at the resonance pole mass $\delta^{(2n)}(M_{2n})=0$. The corresponding widths are extracted analytically:
\begin{equation}
     \Gamma_{2n}=2\lim_{E\to M_{2n}}\sqrt[2n+1]{(-1)^{n-1} (2n)!\frac{M_{2n}-E}{\delta^{(2n)}(E)}}.
\end{equation}
 The residue phase is then unambiguously evaluated as $\theta_{2n}=2\,\delta(M_{2n})-180^\circ$.
 
For genuine, isolated poles, the extracted parameters for the lowest few even derivatives converge ($M_2\approx M_4$ and $\Gamma_2\approx \Gamma_4$). Conversely, the presence of hidden dynamics or nearby singularities---such as subthreshold Adler zeros---breaks this local analytic symmetry, causing the higher-order extractions to explicitly diverge. As shown in Table \ref{Table1}, the canonical vectors $\rho(770)$ and $K^*(892)$ exhibit perfect second-to-fourth derivative (D2/D4) convergence. In stark contrast, the $f_0(500)$ parameters shift systematically, providing a clear, extrapolation-free fingerprint of the non-local dynamical effects dominating the scalar sector. For $K_0^*(700)$, due to its closeness to the threshold, there is no resonant signal even in the first derivative, let alone the higher ones. 

{\bf Fitting the Constructed Data.}---We construct the elastic scattering amplitude $T$ from the phase shifts and search for the pole positions and their residues directly in the complex energy plane (CP). This convention (pole in $E$, not in $s$) is used in the field of nucleon resonances, and we adopt it here because it gives us a unified framework for distinct types of hadrons. 

\begin{figure*}[]
\includegraphics[width=0.32\textwidth]{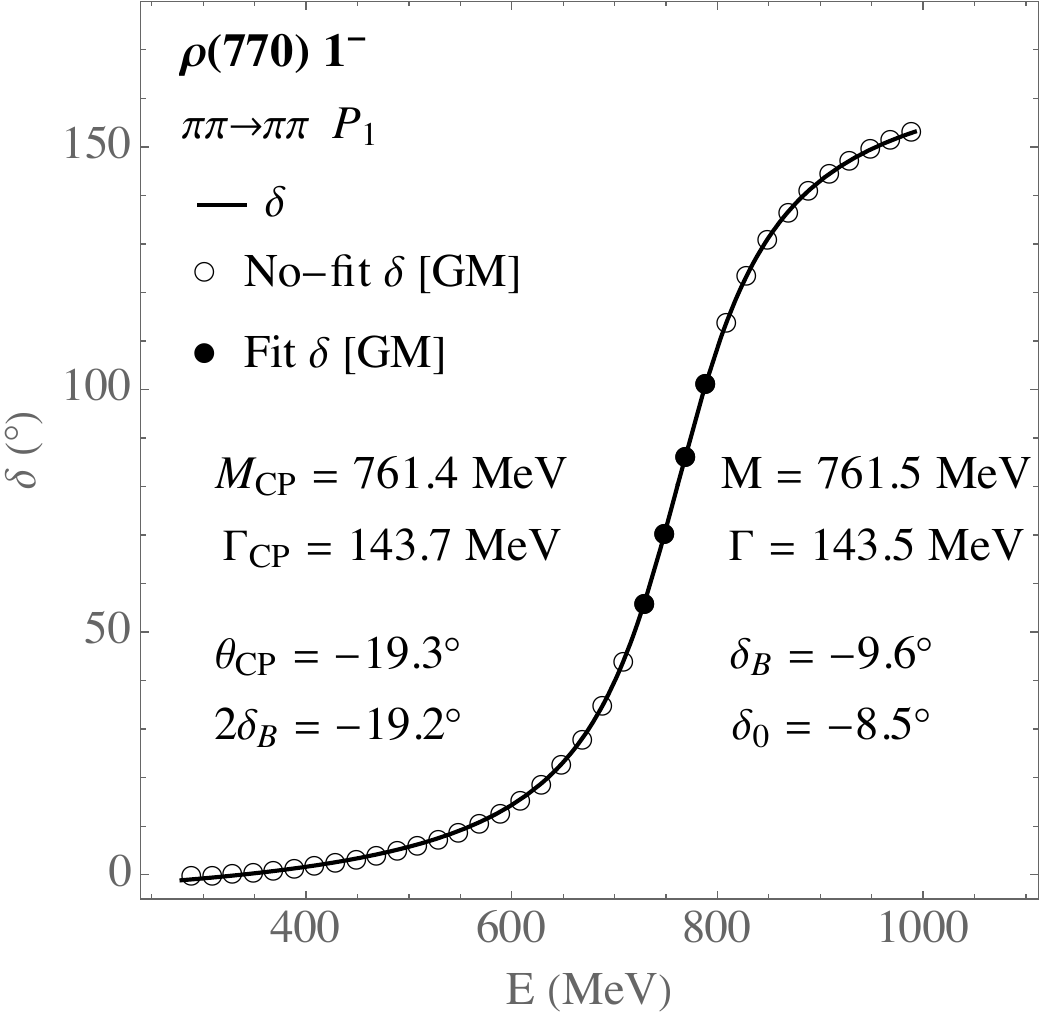}
\includegraphics[width=0.32\textwidth]{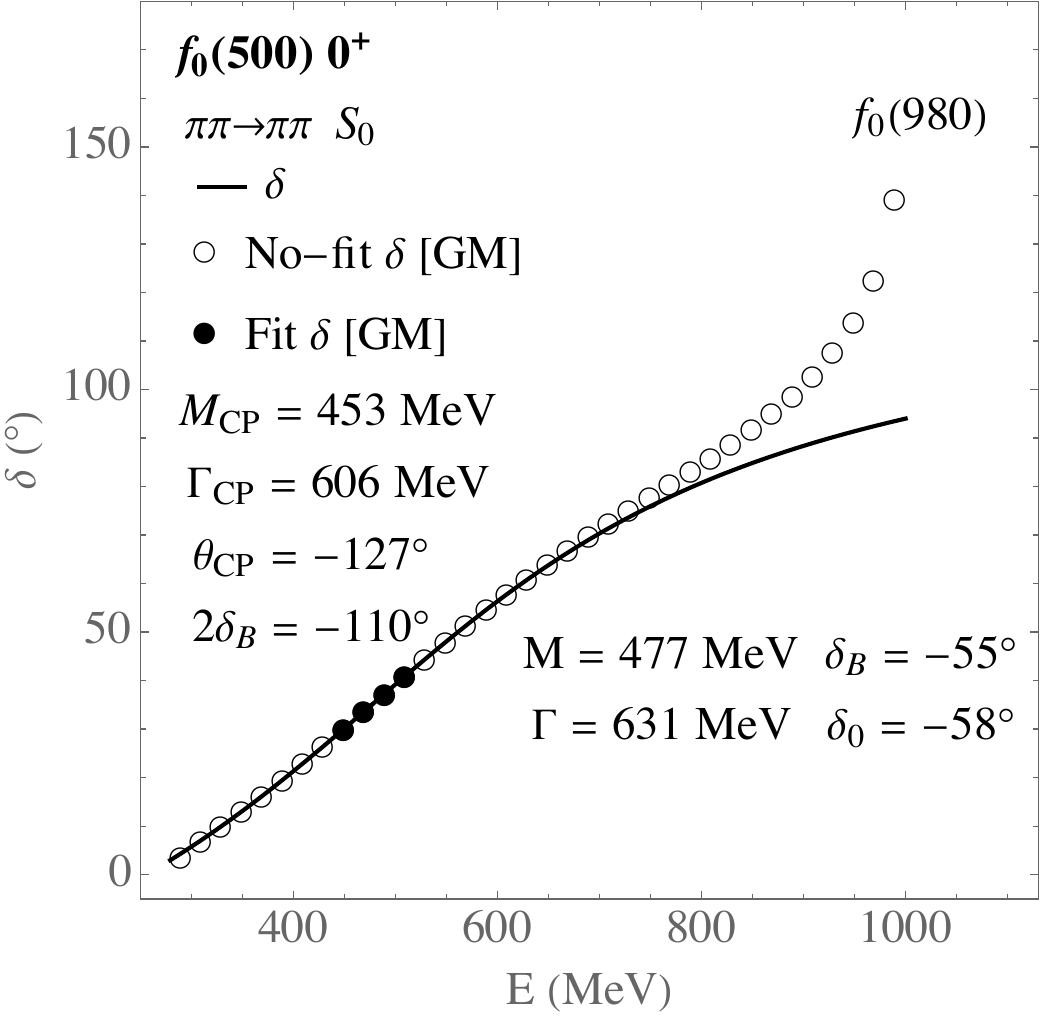}\\
\includegraphics[width=0.32\textwidth]{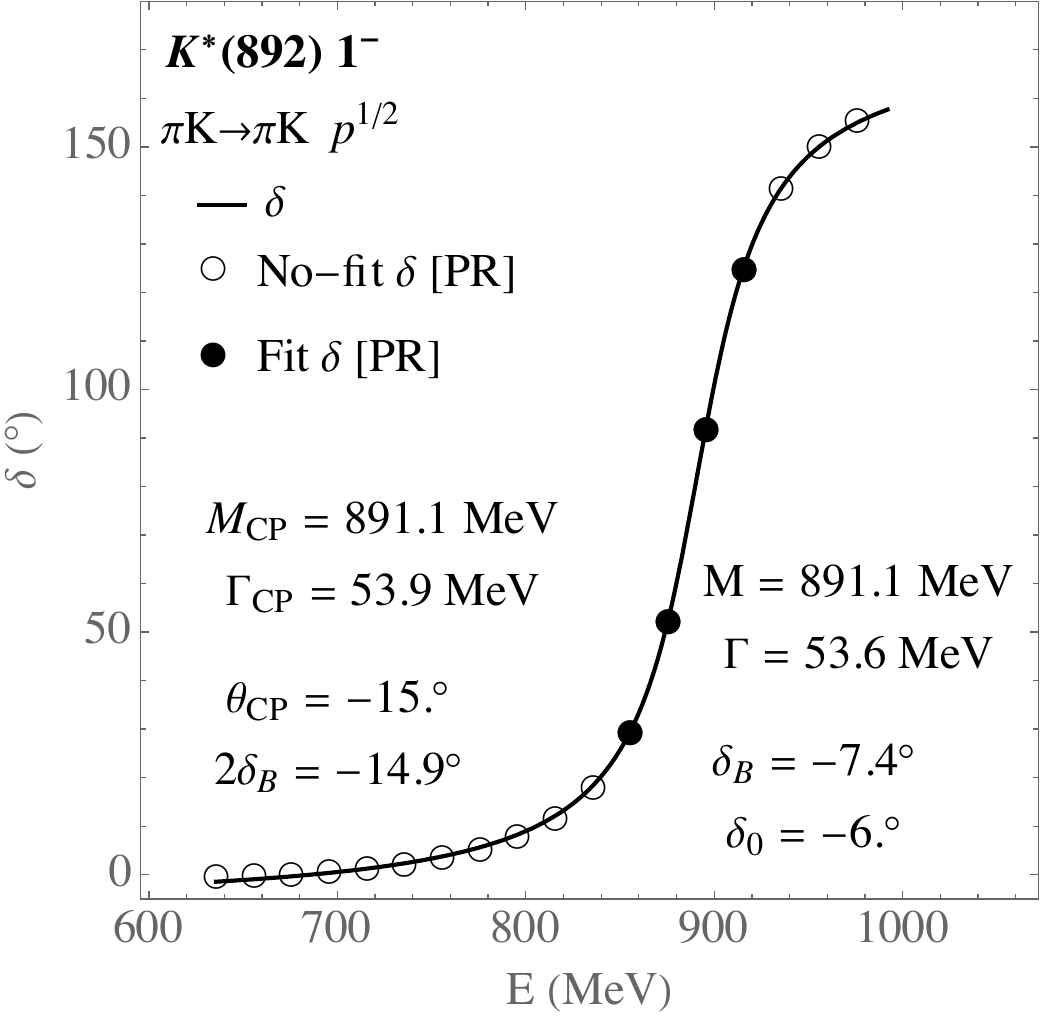}
\includegraphics[width=0.32\textwidth]{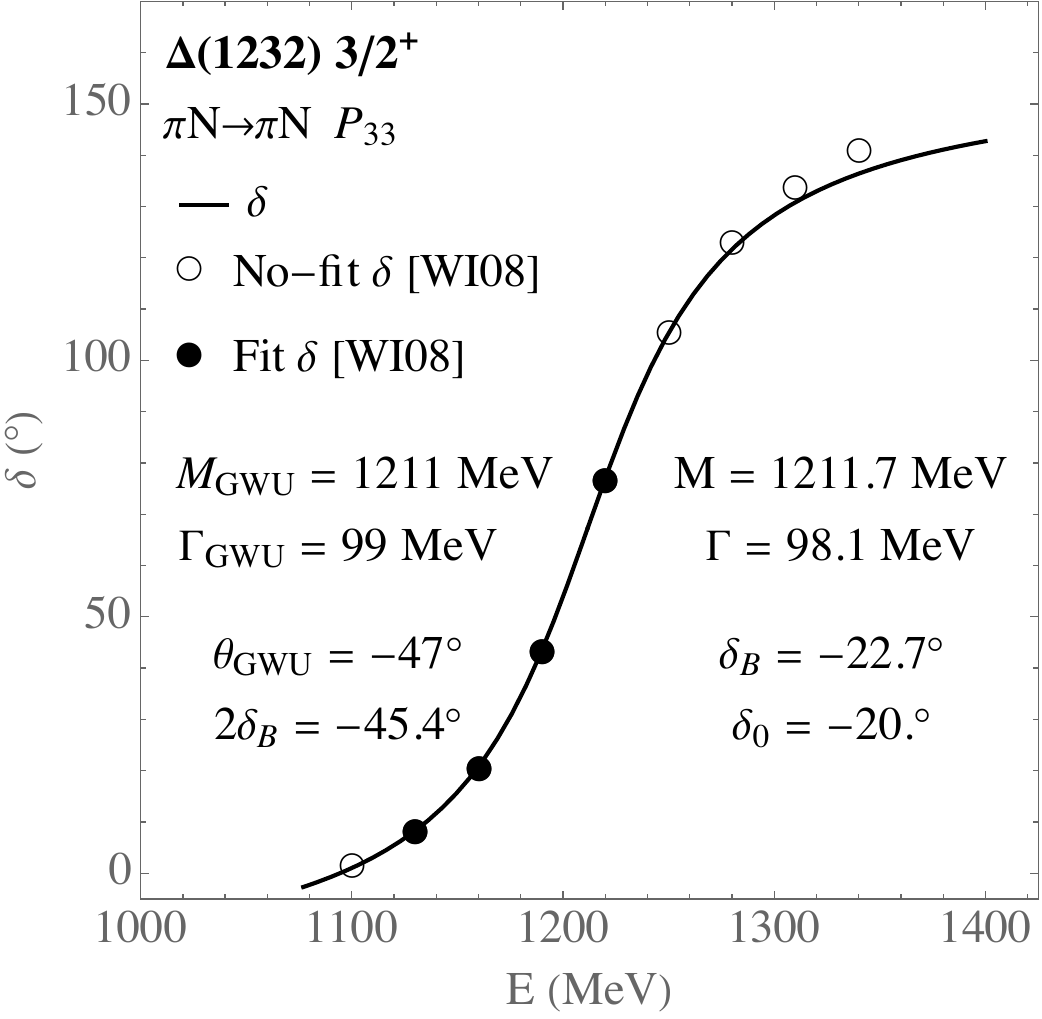}
\includegraphics[width=0.32\textwidth]{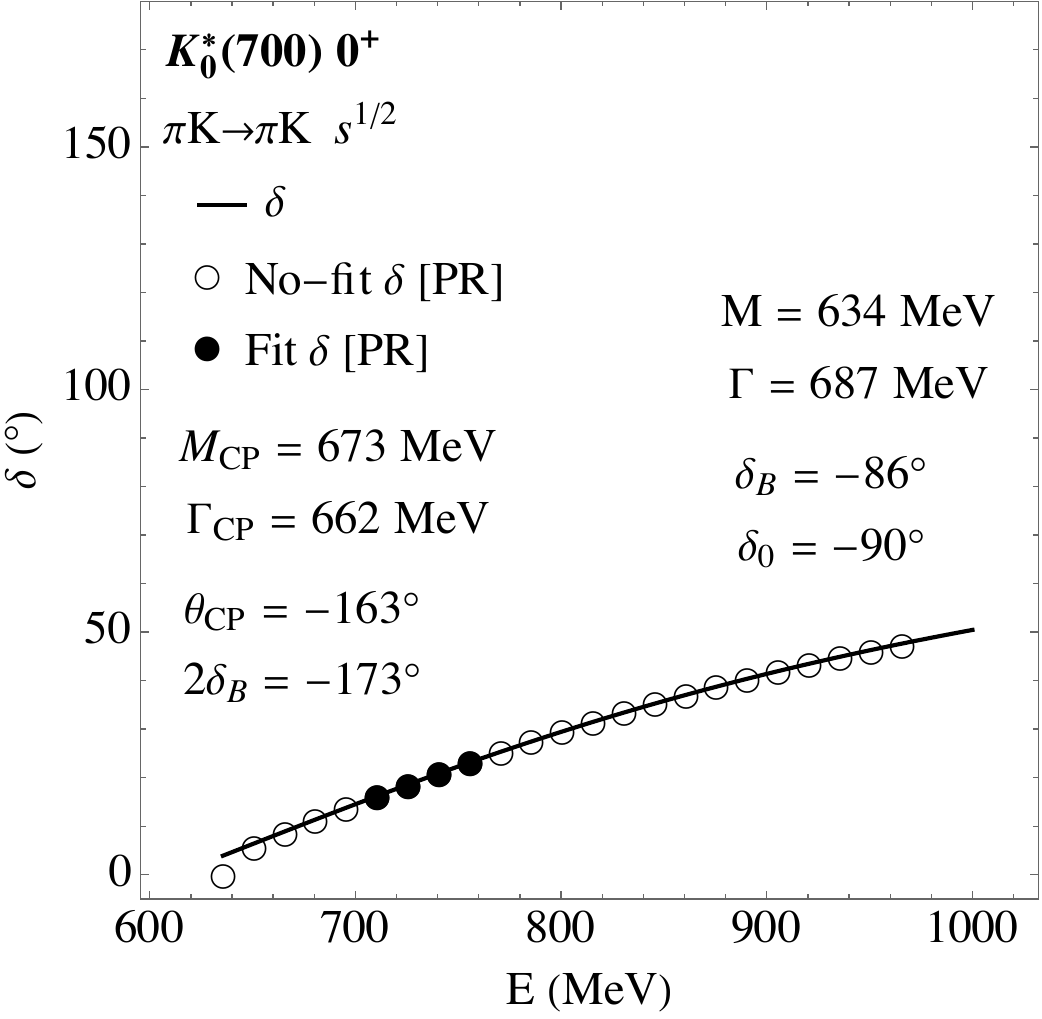}
\caption{Fitting our three-parameter model (OM) from Eq.~\ref{eq:phaseshift} to the phase shifts from Refs.~\cite{GM,PR,Arndt}. Only the four black circle data points at each graph are fitted. For comparison we show the pole parameters directly calculated at the pole position in the complex plane (CP). Since a mathematical model is used to generate the data, the $\chi_R^2$ has no statistically relevant value. Note how $\Delta(1232)$ stands half-way from Breit-Wigner to non-Breit-Wigner resonance and, as a Rosetta stone of a kind, possesses the features of both worlds (nice resonant shape, but also large background and significant residue phase).   
\label{Fig2}}
\end{figure*}

The near-resonance region is isolated by fitting Eq.~(\ref{eq:phaseshift}) to sliding windows of four consecutive theoretical data points, selecting the local parametrization that minimizes $\chi^2$. This procedure ensures a stable extraction of $M$, $\Gamma$, and $\delta_\mathrm{B}$, even for broad resonances where the background variation is non-negligible farther away from the pole. The results of the fit are shown in Fig.~\ref{Fig2}.

Our analysis of the vector sector ($\rho(770)$ and $K^*(892)$) shows nearly perfect agreement between OM fits and CP values. For these states, the extrapolated fitting curve covers the entire elastic range, and the residue phase $\theta$ is accurately predicted by $2\delta_\mathrm{B}$, validating the constant-background approximation even in the presence of a dominant centrifugal barrier.

In the scalar sector ($f_0(500)$ and $K_0^*(700)$), we observe a stable OM fit despite the non-resonant phase shift shapes. The 25~MeV mass discrepancy and the $10^\circ$--$15^\circ$ phase drift from CP values are systematic. This identifies the limit where our geometric model meets chiral dynamics. The Adler zero adds a correction to the unitarity-locked geometric phase.

We note that $f_0(500)$ and $K_0^*(700)$ share an intriguing feature with the relatively narrow $\Delta(1232)$ $\pi N$ elastic resonance: all three are unusually close to their thresholds. Despite being substantially narrower, the $\Delta(1232)$ has a residue phase more than double that of the $\rho(770)$. This strongly suggests that the geometric angle $\delta_0$, and not the width alone, dictates the phase. All our results are collected in Table \ref{Table1}.

\begin{table}[h!]
\caption{The results of fitting our model (OM) to the phase shifts from Refs.~\cite{GM,PR}. We write $2\delta_0$ and $2\delta_\mathrm{B}$ for easier comparison with phase shift $\theta$. The L+P method \cite{Sva14} is used on scattering amplitude constructed from the phase shifts. D2/D4 is a derivative method we propose for distinguishing genuine from dynamical poles (if $D2\neq D4$, there is hidden dynamics). CP results are evaluated at the pole position in the complex plane. PR, GM and GWU are results of the original analyses from Refs.~\cite{PR, GM, Arndt}. HR is from Ref.~\cite{Hof24}, and PDG estimates are from Ref.~\cite{PDG}. \label{Table1} }
\resizebox{\columnwidth}{!}
{%
\begin{tabular}{llllll} 
\hline
\hline
State & Src. & $M$ (MeV) & $\Gamma$ (MeV) & $2\delta_\mathrm{B}\,|\,\theta$  ($^\circ$) & $2\delta_0$ ($^\circ$) \\
 \hline
 $K^*(892)$ & OM  & $891.1$ & $53.6$ & $-14.9$ & $-12.0$ \\
            & L+P & $891.0$  & $53.7$   & $-15.0$ & $-12.0$ \\
            & D2 & $891.1$ & $53.9$  & $-14.8$ & $-12.0$ \\
            & D4 & $891.1$ & $53.9$  & $-14.9$ & $-12.0$ \\            
            & CP  & $891.1$ & $53.9$ & $-15.0$ & $-12.0$\\
            & PR  & $890\pm2$ & $51.6\pm2.4$ & $-15.2\pm1.2$ & $-11.6$\\
            & PDG & $890\pm14$ & $52\pm12$ & $---$ & $-11.7\pm2.8$ \\
\hline 
 $\rho(770)$ & OM  & $761.5$ & $143.5$ & $-19.2$ & $-16.9$ \\
             & L+P & $761.3$  &  $143.3$   & $-19.4$ & $-16.9$ \\
             & D2 & $761.5$ & $143.7$  & $-19.2$ & $-16.9$ \\
             & D4 & $761.4$ & $143.7$  & $-19.2$ & $-16.9$ \\
             & CP  & $761.4$ & $143.7$ & $-19.3$ & $-16.9$\\
             & GM  & $763.7^{+1.7}_{-1.5}$ & $146.4^{+2}_{-2.2}$ & $-18.6\pm1.0$ & $-17.2$\\
             & HR  & $760.6\pm0.8$ & $142.4\pm0.9$ & $-19.0\pm0.6$ & $-16.8$ \\
             & PDG & $763\pm2$ & $145\pm3$ & $---$& $-17.0\pm0.4$ \\
 \hline 
 $\Delta(1232)$ & OM  & $1212$ & $98$ & $-45$ & $-40$ \\
                & L+P &  $1211$ & $98$   & $-46$   & $-40$ \\
                & GWU & $1211$ & $99$   & $-47$   & $-41$\\
                & PDG & $1210\pm1$ & $100\pm2$ & $-46^{+1}_{-2}$& $-41\pm2$ \\
 \hline 
 $f_0(500)$ & OM  & $474$ & $629$  & $-110$ & $-116$ \\
            & L+P & $485$ & $618$  & $-107$ & $-113$ \\
            & D2 & $477$ & $627$  & $-110$ & $-116$ \\
            & D4 & $461$ & $578$  & $-116$ & $-116$ \\
            & CP  & $453$ & $606$  & $-127$ & $-120$\\
            & GM  & $457^{+14}_{-13}$ & $558^{+22}_{-14}$ & $-118\pm12$ & $-115$\\
            & HR  & $449\pm6$ & $550\pm24$ & $-124\pm10$& $-117$\\
            & PDG & $400\,$--$550$ & $400\,$--$700$ & $---$& $-109\pm25$ \\
 \hline 
 $K_0^*(700)$ & OM  & $634$ & $687$  & $-173$ & $-181$ \\
              & L+P &  $637$  & $674$ & $-171$ & $-180$ \\
              & D2/4 &  \multicolumn{4}{c}{Not  seen} \\
              & CP  & $673$ & $662$  & $-163$ & $-167$\\
              & PR  & $648\pm7$ & $560\pm32$ & $-180\pm8$& $-175$\\
              & PDG & $630\,$--$730$ & $520\,$--$680$ & $---$ & $-163\pm19$ \\
\hline\hline
\end{tabular}
}
\end{table}

To further test our results, we apply the L+P method, which has successfully been used in nucleon spectroscopy for extractions of the pole and residue properties \cite{Sva14}. In addition to Laurent part handling the poles, it uses Pietarinen expansion with conformal variable taking care of channel openings. The L+P results are fully consistent with ours, which suggests that Adler zeros are effectively subsumed within the local behavior of the data. In addition, it means that L+P should be modified for scalar mesons to include Adler zeros in order to  precisely extract properties of the dynamical poles. 

Indeed, when we compare the fits of our model (OM) and L+P, and the relevant results from the literature \cite{GM,PR,Hof24}, which we had to translate to the nucleon resonance convention using Eq.~(\ref{eq:Convention}), the agreement in the residue phases seems striking. 

The systematic drift observed in the scalar sector is not a failure of our geometric model but a clear identification of dynamical contributions. While the pole-threshold geometry provides ``skeleton'' of the residue phase (accounting for over 85\% of its value), the Adler zero, required by chiral symmetry, adds a well-defined correction. This confirms our previous thesis \cite{Cec26, Cec26B} that residue phases are largely fixed by unitarity, with the remaining physics residing in the precise pole placement and subtle dynamical shifts.

\begin{figure}[h]
\includegraphics[width=0.45\textwidth]{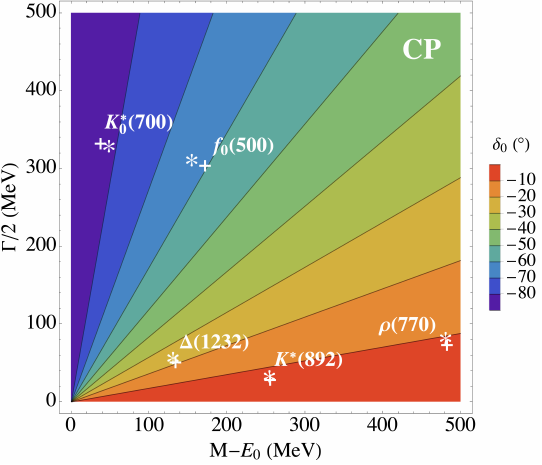}
\caption{
Phase rotation plot showcasing the geometric consistency across different hadronic sectors. Each ``+" denotes the pole position ($M-E_0$,$\Gamma/2$) extracted from established dispersive analyses (CP). The ``$*$" indicates the position if the geometric phase $\delta_0$ were replaced by $\theta/2$. The proximity of ``+" and ``$*$" symbols across $\pi \pi$, $\pi K$, and $\pi N$ sectors confirms that the residue phase is fundamentally driven by the threshold geometry.
\label{Fig3}}
\end{figure}

The CP results further confirm our main assumption---that the main driving term of the residue phase $\theta$ is our geometric angle $2\delta_0$. To see that more clearly, in Fig.~\ref{Fig3} we show a combined plot of all the CP results (designated with the cross sign), and how much would pole position rotate if instead of $\delta_0$ there was $\theta/2$ (designated by the asterisk). The difference is surprisingly small even for the $f_0(500)$ and $K_0^*(700)$. 

The geometric origin of the residue phase is robust, but not completely rigid. As formally demonstrated by our D2/D4 diagnostics (Table \ref{Table1}), the subtle deviations observed in the scalar sector act as a direct quantitative measure of the underlying chiral dynamics.

%\section{Conclusion}

{\bf Conclusion.}---The fact that the simple pole plus constant background fits almost perfectly various theoretical elastic phase shifts for $\pi \pi$, $\pi K$, and $\pi N$ scattering \cite{PR,Arndt,GM} tells us that this is a reasonable approximation quite far in the complex plane. That means the constant background $\delta_\mathrm{B}$ can do the heavy lifting of setting both the scattering amplitude to zero at the threshold (where the total phase shift $\delta$ will be zero), setting its imaginary part to unity at the Breit-Wigner peak position (where $\delta$ is $90^\circ$), and setting the residue phase to $\theta=2\delta_\mathrm{B}$ at the pole position. Consequently, the value of the background phase shift is to a significant extent given by the geometric phase $\delta_0$, defined as the angle between the pole position and the real energy axis seen from the threshold. This has been suggested by imposing the S-matrix unitarity on the elastic scattering amplitudes from Refs.~\cite{Cec26,Cec26B}. There, it was shown that such amplitude is in fact a mathematical identity valid when there are no additional singularities or zeros between the threshold and the pole. Indeed, we have seen the strongest departure of  $\delta_\mathrm{B}$ from $\delta_0$ for the spin-0 partial waves where there is Adler zero present in the proximity of the threshold. Finally, our main hypothesis is once more confirmed by naturally expanding the theoretical phase shifts from the literature into the complex plane, and finding that $\theta/2$ and $\delta_0$ are almost indistinguishable. The anomalous nature of the $f_0(500)$ and $K_0^*(700)$ states is thus naturally demystified: their extreme pole locations mandate a maximal geometric angle $\delta_0$, fundamentally detaching the Breit-Wigner peak from the pole mass while dictating a nearly constant background phase across the entire elastic domain.

{\bf Data Availability.}---The data that support the findings of this study are openly available in Zenodo \cite{ZenodoData}.

\end{document}